\documentclass[final]{ias2}

\usepackage{graphicx} 
\usepackage{multirow}
\usepackage{array} 

\usepackage{hyperref} 
\begin{document}

\markboth{Predictions From High Scale Mixing Unification Hypothesis}{Rahul Srivastava}

 %%%%%%%%%%%%%%%%%%%%%%%%%%%%%%%%%%%%%%%%%%%%%%%%%%%%%%%%%%%%%%%%%%%%%%%%%%%%%%%%%%%%%%%%%%%%%%%%%%%%%%%%%%%%%%%%%%%%%%%%%%%%%%%%%%%%%%%%%%%%%%%%%%%%%%%%%%%%%%%%%%%%%%%%%%%%%%%%%%%%%%
 
\title{Predictions From High Scale Mixing Unification \\ Hypothesis}

\author[imsc]{Rahul Srivastava} 
\email{rahuls@imsc.res.in}
\address[sin]{The Institute of Mathematical Science, Chennai, India - 600113}

\begin{abstract}
 Starting with 'High Scale Mixing Unification' hypothesis, we investigate the renormalization group evolution of mixing
parameters and masses for both Dirac and Majorana type neutrinos. Following this hypothesis, the PMNS mixing
parameters are taken to be identical to the CKM ones at a unifying high scale. Then, they are evolved to a low scale using
MSSM renormalization-group equations. For both type of neutrinos, the renormalization group evolution “naturally”
results in a non-zero and small value of leptonic mixing angle $\theta_{13}$. One of the important predictions of this analysis is that, 
in both cases, the mixing angle $\theta_{23}$ turns out to be non-maximal for most of the parameter range. We also elaborate on the important differences
between Dirac and Majorana neutrinos within our framework and how to experimentally distinguish between the two scenarios. Furthermore, for both cases, we also derive
constraints on the allowed parameter range for the SUSY breaking and unification scales, for which this hypothesis works.
The results can be tested by present and future experiments.
\end{abstract}

\keywords{Neutrino: oscillation, Neutrino: Majorana, Neutrino: Dirac, Supersymmetry: symmetry breaking, PMNS matrix, Mixing angles, Renormalization group}

\pacs{14.60.Pq, 11.10.Hi, 11.30.Hv, 12.15.Lk}
 
\maketitle

%%%%%%%%%%%%%%%%%%%%%%%%%%%%%%%%%%%%%%%%%%%%%%%%%%%%%%%%%%%%%%%%%%%%%%%%%%%%%%%%%%%%%%%%%%%%%%%%%%%%%%%%%%%%%%%%%%%%%%%%%%%%%%%%%%%%%%%%%%%%%%%%%%%%%%%%%%%%%%%%%%%%%%%%%

\section{Introduction}

%%%%%%%%%%%%%%%%%%%%%%%%%%%%%%%%%%%%%%%%%%%%%%%%%%%%%%%%%%%%%%%%%%%%%%%%%%%%%%%%%%%%%%%%%%%%%%%%%%%%%%%%%%%%%%%%%%%%%%%%%%%%%%%%%%%%%%%%%%%%%%%%%%%%%%%%%%%%%%%%%%%%%%%%%%%%%%%

   Neutrinos are probably the most mysterious and ill understood of all known particles. In past neutrinos have thrown up quite a few surprises and they still keep on surprising us.
 The recent measurements have conclusively shown that the neutrino mixing angle $\theta_{13} \neq 0$ \cite{Abe:2011sj,Adamson:2011qu,Abe:2012tg,Ahn:2012nd,An:2012eh}.   
 Measurement of $\theta_{13}$ was long awaited as it provides crucial test of several candidate PMNS mixing ansatzes like Tri-Bi-Maximal (TBM) mixing ansatz which predicted $\theta_{13} = 0$.  
 Is there a ``natural'' way of understanding non-zero and ``relatively large'' $\theta_{13}$?    
 In this work we  explore one such possibility namely the High Scale Mixing Unification (HSMU) of CKM aqnd PMNS mixing matrices.

 Unification of seemingly unrelated phenomenon is an old and quite fruitful notion. In past it has lead to much advancement in our understanding e.g. Electro-Magnetism, Electro-Weak force etc.
  Currently lot of research has been devoted to study and construct models for unification of the fundamental forces.  Such Grand Unified Theories (GUTs) attempt to unify the three fundamental forces namely electromagnetism, weak and strong forces. One key aspect of GUTs is the unification of gauge couplings.  Another key ingredient of GUTs is that the quarks and leptons are in same multiplet of the GUT gauge group. Hence in GUTs the flavor structure of quarks and leptons is not totally disconnected. In view of this  we try to explore the interesting possibility of ``High Scale'' unification of CKM and PMNS mixing parameters.

One may wonder how is this possible at all? We have measured the CKM mixing parameters quite well \cite{Agashe:2014kda} and now have a fairly good idea about the PMNS mixing angles as well \cite{GonzalezGarcia:2012sz} and numerically they are very different from each other. The quark mixing matrix is almost diagonal with small off diagonal elements resulting in small mixing angles, whereas the neutrino mixing angles are relatively large.

 The important point to note here is that the unification of CKM and PMNS mixing parameters is expected to occur at high scales e.g GUT scale. Hence, one needs to use RG equations to obtain values of the quark and lepton mixing parameters at the electroweak scale before comparing them with the experimental values. Now, it is known that since the quark masses are hierarchical in nature, therefore the quark mixing angles do not change much in SM or even in MSSM RG running. What about neutrino mixing angles? In this work we look in details about the possibility of large radiative magnification of PMNS mixing angles.

Before presenting our results let us briefly discuss how large the radiative magnification should be to make HSMU hypothesis consistent with present low scale neutrino oscillation data.  The high scale mixing unification implies that CKM angles = PMNS angles. 
 More specifically, for unification at some ``High Scale'', say GUT we should have
\begin{eqnarray}     
\theta^{0,q}_{12}=\theta^0_{12}=13.02^\circ, \quad \theta^{0,q}_{13} = \theta^0_{13}= 0.17^\circ, \quad \theta^{0,q}_{23} = \theta^0_{23}= 2.03^\circ
\label{uniang}
\end{eqnarray}
where we have used a superscript ``0'' to distinguish these high scale values from the values obtained at low scale.
 Since, the CKM mixing angles do not change much,  large radiative magnification of PMNS angles is required for it to be consistent with present experimental measurements. This implies 
\begin{eqnarray}     
&&\theta^{0}_{12}=13.02^\circ \rightarrow \theta_{12}=33.36^\circ, \quad \theta^{0}_{13}= 0.17^\circ  \rightarrow \theta_{13}= 8.66, \nonumber \\
&& \qquad \qquad \qquad \theta^{0}_{23} =  2.03^\circ \rightarrow    \theta^0_{23}= 40.0 \oplus 50.4
\end{eqnarray} 
 Such large radiative magnification can be realized within Minimal Supersymmetric Standard Model (MSSM) \cite{Mohapatra:2003tw, Mohapatra:2004vw, Mohapatra:2005gs, Mohapatra:2005pw, Agarwalla:2006dj,Abbas:2013uqh, Abbas:2014ala, HSMU_phase} and in this work we intend to investigate the predictions obtained in such a scenario. This contribution is based on \cite{Abbas:2013uqh, Abbas:2014ala, HSMU_phase} and the interested reader is referred to them for further details.

  The plan of this proceeding is as follows. In Section \ref{sec2} starting with HSMU hypothesis we present our results for the case of Majorana neutrinos assuming no CP violation in lepton sector. In Section \ref{sec3} we discuss the results of HSMU hypothesis for Dirac neutrinos. In Section \ref{sec4} we derive constraints on the unification scale, SUSY breaking scale and $\tan \beta$ for which our analysis works. In Section \ref{sec5} we study the effects of non-zero phases on the results obtained in earlier sections. In Section \ref{sec6} we briefly summarize our main results and the various tests for HSMU hypothesis. We conclude in Section \ref{sec7}.

%%%%%%%%%%%%%%%%%%%%%%%%%%%%%%%%%%%%%%%%%%%%%%%%%%%%%%%%%%%%%%%%%%%%%%%%%%%%%%%%%%%%%%%%%%%%%%%%%%%%%%%%%%%%%%%%%%%%%%%%%%%%%%%%%%%%%%%%%%%%%%%%%%%%%%%%%%%%%%%%%%%%%%%%%

\section{Majorana case}
\label{sec2}

%%%%%%%%%%%%%%%%%%%%%%%%%%%%%%%%%%%%%%%%%%%%%%%%%%%%%%%%%%%%%%%%%%%%%%%%%%%%%%%%%%%%%%%%%%%%%%%%%%%%%%%%%%%%%%%%%%%%%%%%%%%%%%%%%%%%%%%%%%%%%%%%%%%%%%%%%%%%%%%%%%%%%%%%%%%%

  In this section we discuss the case of Majorana neutrinos. We take a model independent approach and assume HSMU at some ``High Scale''. The details of the ``High Scale'' theory are not needed for our analysis. Below the high scale we assume MSSM with Type-I seesaw mechanism. The right handed neutrinos are integrated out below their mass threshold and below seesaw scale we have effective dimension five neutrino mass operator.

 For testing HSMU one needs to run down the masses and mixing parameters from high scale to low scale ($M_Z$). The RG running between high scale and seesaw scale is done using standard MSSM RG equations within framework of Type-I seesaw mechanism. 
 Below seesaw scale the RG running is done with dim-5 operator added to MSSM. After SUSY breaking scale we do the RG running with dim-5 operator added to SM. 
 The RG equation for the neutrino masses, PMNS mixing angles and phases can be found in \cite{Antusch:2002ek, Antusch:2003kp, Antusch:2005gp, Casas:1999tg}.

The natural ``High Scale'' for HSMU hypothesis is the scale of Grand Unified Theories (GUTs). Therefore in this section we will assume that HSMU is realized at GUT scale i.e. $2 \times 10^{16}$ GeV.  Sensitivity to choice of high scale will be discussed in Section \ref{sec4}. Also, the HSMU hypothesis is  realized for varied range of seesaw scales but for sake of definiteness we will choose typical seesaw scale of $10^{12}$ GeV. In this section we will take the SUSY breaking scale of 5 TeV and the dependence on choice of SUSY breaking scale will be discussed in Section \ref{sec4}. Since larger values of $\tan \beta$ lead to enhanced magnification therefore in this section we will take $\tan\beta = 55$ and the dependence of our analysis on  $\tan \beta$ will be discussed in Section \ref{sec4}. 

Before presenting the results of our analysis we will like to point out that the HSMU hypothesis can only be implemented for the case of normal hierarchy. In case of inverted hierarchy the large radiative magnification of $\theta_{23}$ cannot be achieved \cite{Abbas:2014ala}. Moreover, for large radiative magnification of the mixing angles we require the neutrinos to be quasi degenerate. Also, for sake of simplicity, in this section we will only discuss the case of no CP violation in leptonic sector and will take the  Dirac as well as Majorana phases to be zero. The  CP violating scenario and effect of phases on HSMU will be discussed in Section \ref{sec5}.

 The HSMU hypothesis is implemented in two steps. We start from  known values of gauge couplings, quark mixing angles, masses of quarks and charged leptons at the low scale ($M_Z$) and use RG equations to obtain the corresponding values at the high scale.
 At the high scale we assume HSMU i.e. we take the neutrino mixing angles same as the quark mixing angles at this scale. The neutrino masses at high scale are treated as unknown parameters. These three parameters are determined by the requirement that the low energy values of the oscillation parameters i.e.  $\Delta m_{12}^2, \Delta m_{23}^2, \theta_{12}, \theta_{23}$ and $\theta_{13}$ agree with their present experimental ranges.

\begin{figure}[ht]
 \vspace{0.2cm}
\begin{center}
\includegraphics[width=0.4\columnwidth]{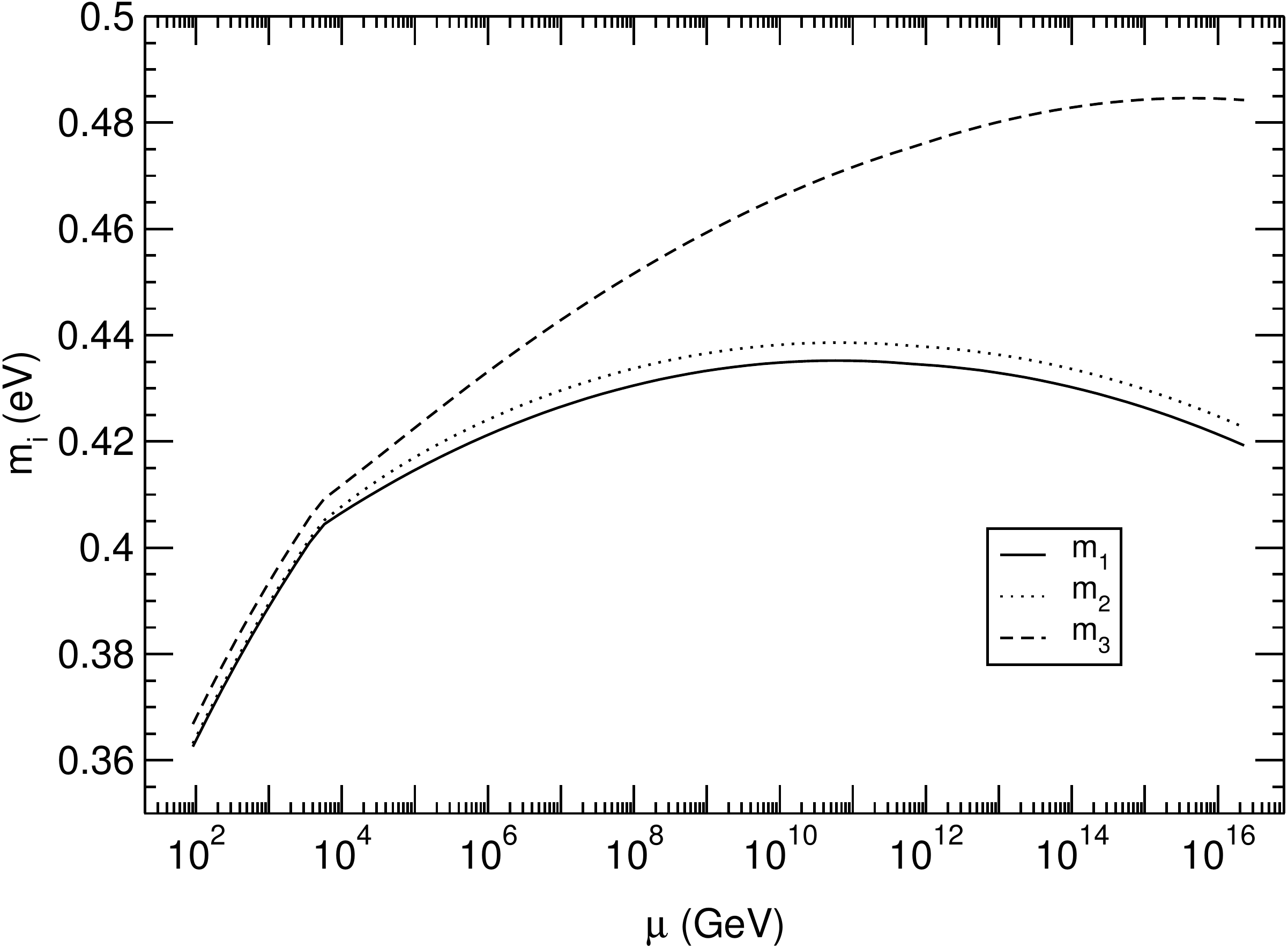}  \hspace{0.1cm}
\includegraphics [width = 0.4\columnwidth]{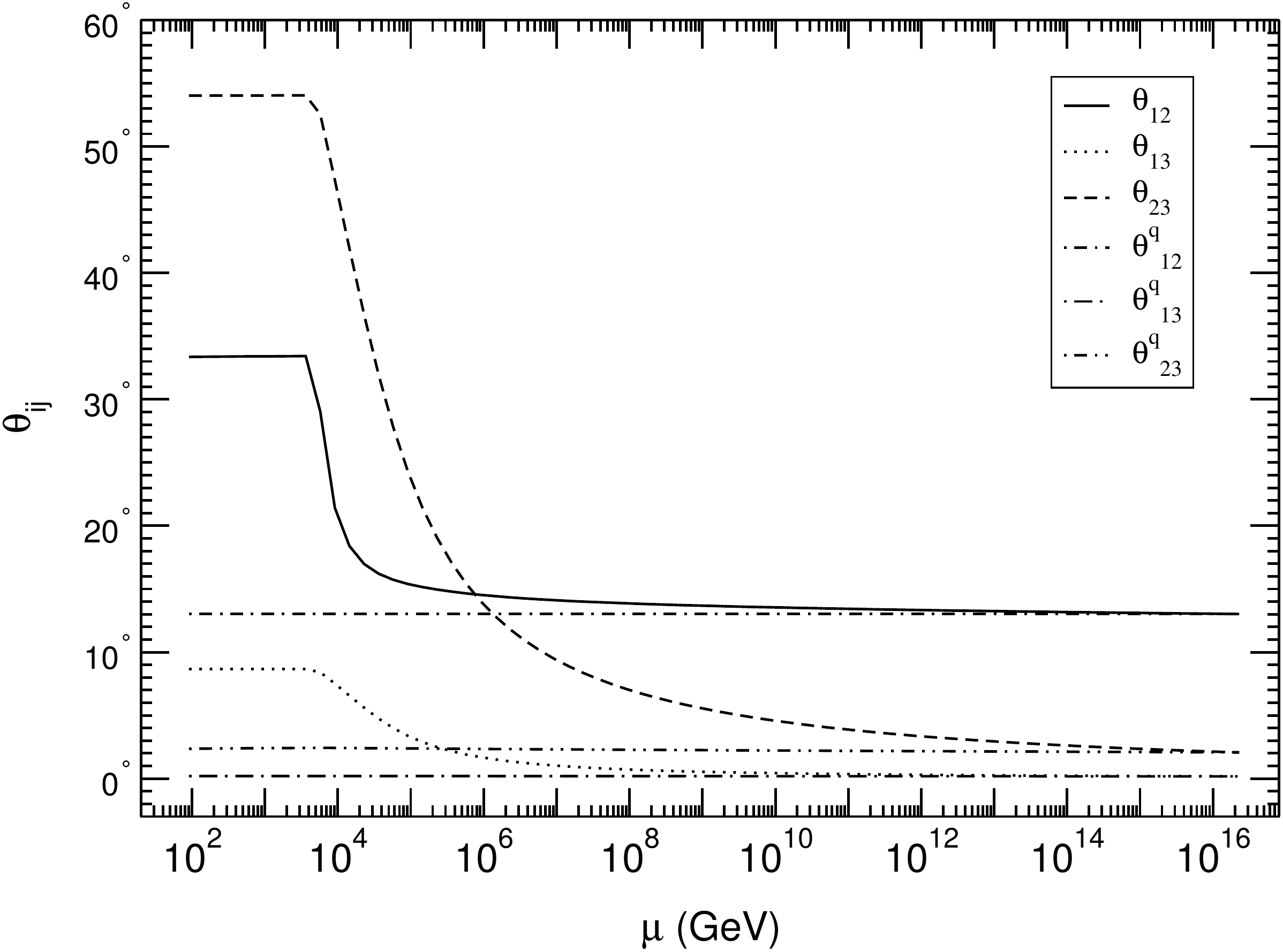}
\caption{The RG evolution of neutrino masses as well as  CKM and PMNS mixing angles with respect to RG scale $(\mu)$.}
\label{fig1}
\end{center}
\end{figure}

The RG running of neutrino masses and PMNS mixing angles are shown in Fig \ref{fig1}.

 As is clear from Fig \ref{fig1}, all masses decrease from unification scale to $M_Z$. They acquire nearly degenerate masses at $M_Z$. Also the dominant contribution to the RG running of both CKM and PMNS mixing angles can be approximately given by   

 \begin{eqnarray}
 \frac{d \theta_{12}}{dt}  & \propto &  \frac{m^2}{\Delta m^2_{21}}; \qquad \frac{d \theta_{13}}{dt}, \, \frac{d \theta_{23}}{dt}  \, \propto \, \frac{m^2}{\Delta m^2_{32}} 
 \label{angrg}
 \end{eqnarray}

 where $t=\ln(\mu)$ and $\mu$ is the renormalization scale. From (\ref{angrg}) it is clear that owing to the hierarchical nature of quark masses, the RG running of CKM mixing angles is almost negligible.  Due to quasi-degenerate neutrino masses large radiative magnification occurs for the PMNS mixing angles.

 Since the RG evolution of the PMNS parameters and neutrino masses are correlated with each other, one can obtain several important constraints on the low scale observables. In Fig \ref{fig3} we show the correlation between the mixing angles $\theta_{13}$ and $\theta_{23}$.
 
 \begin{figure}[ht]
  \vspace{0.5cm}
\begin{center}
\includegraphics[width=0.4\columnwidth]{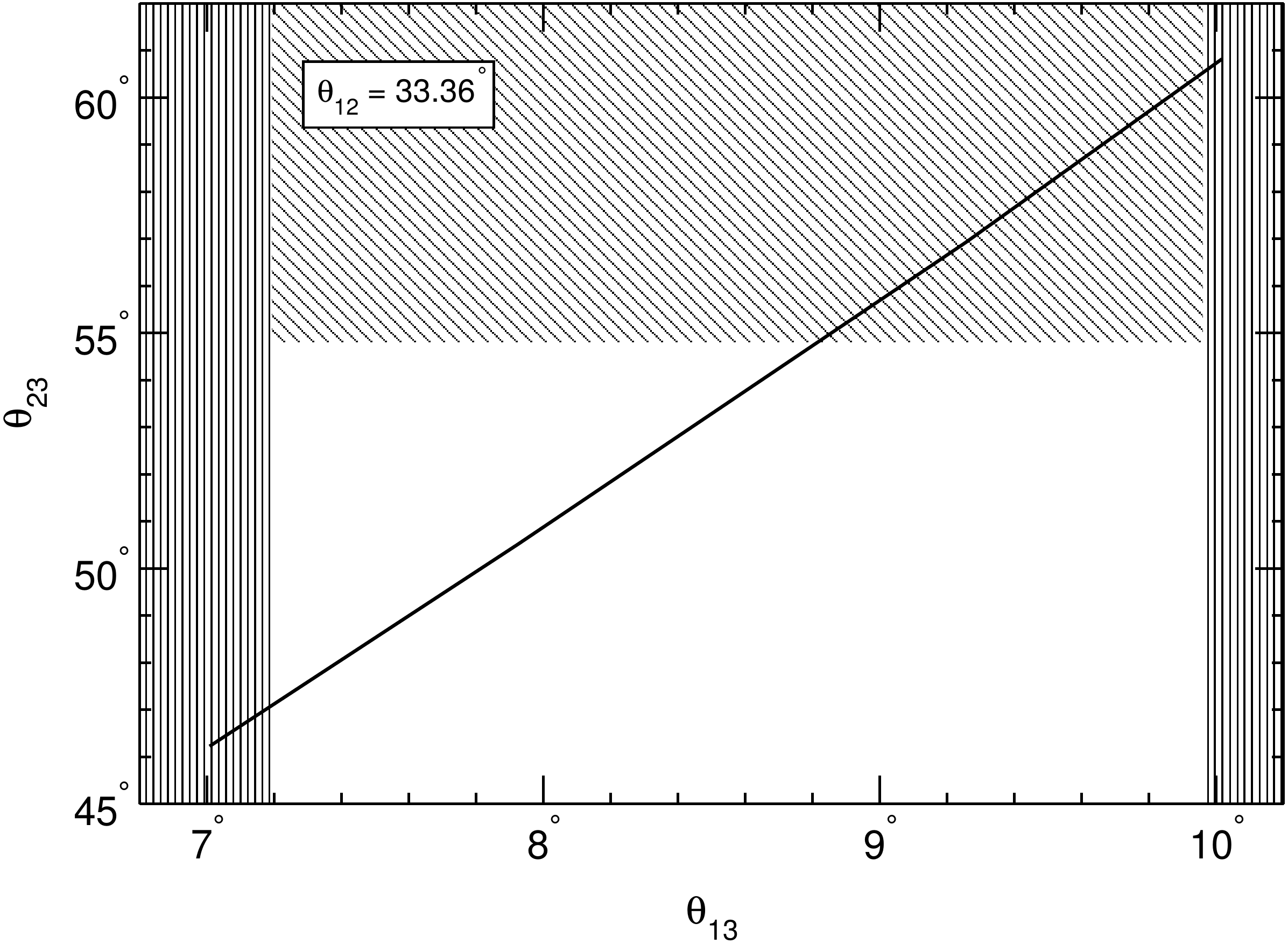}
\caption{The variation of $\theta_{23}$ with respect to $\theta_{13}$. For plotting this figure we have kept all other oscillation parameters to be at their best-fit values. 
         The vertically shaded regions lie outside the 3-$\sigma$ range of $\theta_{13}$ whereas the horizontally shaded 
         one lies outside 3-$\sigma$ range of $\theta_{23}$ \cite{GonzalezGarcia:2012sz}.}
\label{fig3}
\end{center}
\end{figure}

 As is clear from Fig. \ref{fig3}, the mixing angle $\theta_{23}$ is non maximal  i.e. $\theta_{23} > 45^\circ$ and lies in second octant for the  whole  3-$\sigma$ range of $\theta_{13}$.  This prediction is easily  testable in the current and future experimets, like INO, T2K, NO$\nu$A, LBNE, Hyper-K, PINGU \cite{Abe:2011ks,Patterson:2012zs,Adams:2013qkq,Ge:2013ffa,Kearns:2013lea,Athar:2006yb}.

  Also, the mean mass $m= \frac{1}{3}(m_1 + m_2 + m_3)$ lie in the range of ($ \sim 0.34 - 0.38 $) eV. Since we have assumed no CP violation in lepton sector this means that ``Effective Majorana mass'' $m_{\beta\beta} \equiv  \Big[\sum_i U_{ei}^2 \; m_i \Big]$ and ``Averaged electron neutrino mass''  $m_\beta \equiv \Big[\sum_i |U_{ei}|^2 \; m_i^2 \Big]^{1/2}$ are approximately same as mean mass. The present limits from neutrinoless double beta decays experiments are  $\longrightarrow (0.14 - 0.38)$ eV on $<m_{\beta\beta}>$ from EXO-200 experiment \cite{Auger:2012ar}. The limits on $<m_\beta>$ are  $\longrightarrow (< 2)$ eV on $<m_{\beta}>$ from MAINZ, TROITSK experiments \cite{Kraus:2004zw,Aseev:2011dq} with the KATRIN reach of  $\longrightarrow (0.2)$ eV \cite{Drexlin:2013lha}. Thus the predicted values from HSMU hypothesis are  within reach of present experiments and will serve as important tests of HSMU hypothesis.

%%%%%%%%%%%%%%%%%%%%%%%%%%%%%%%%%%%%%%%%%%%%%%%%%%%%%%%%%%%%%%%%%%%%%%%%%%%%%%%%%%%%%%%%%%%%%%%%%%%%%%%%%%%%%%%%%%%%%%%%%%%%%%%%%%%%%%%%%%%%%%%%%%%%%%%%%%%%%%%

\section{Dirac Case}
   \label{sec3}

%%%%%%%%%%%%%%%%%%%%%%%%%%%%%%%%%%%%%%%%%%%%%%%%%%%%%%%%%%%%%%%%%%%%%%%%%%%%%%%%%%%%%%%%%%%%%%%%%%%%%%%%%%%%%%%%%%%%%%%%%%%%%%%%%%%%%%%%%%%%%%%%%%%%%%%%%%%%%%%%%%%%%%%%%%%%

 One of the most important open questions in neutrino physics is whether neutrinos are Dirac or Majorana particles.  Answering this question is essential to find the underlying theory of neutrino masses and mixing. From theoretical perspective, the smallness of mass for Majorana neutrinos is elegantly explained by the sea-saw mechanism. However, even for the Dirac neutrinos there exist a number of appealing models which can explain the smallness of neutrino masses \cite{Ma:2014qra, Ma:2015raa, Mohapatra:1986bd,ArkaniHamed:2000bq,Borzumati:2000mc,Kitano:2002px,Abel:2004tt}. Therefore, current understanding is that Dirac neutrinos are as plausible as Majorana ones.  Neutrinoless double beta decay experiments can potentially resolve this issue but so far they have not seen any signal \cite{Agostini:2013mzu,Auger:2012ar,Gando:2012zm,Alessandria:2011rc}. Therefore it is instructive to see 
 if HSMU can be implemented for Dirac neutrinos as well.

 Here we like to remark that HSMU hypothesis is more natural for Dirac neutrinos than Majorana neutrinos. If neutrinos are Majorana particles then the PMNS-matrix has 6 independent parameters i.e. 3-mixing angles, 1-Dirac phase and 2-Majorana phases. 
  On the other hand CKM-matrix has only 4 independent parameters: 3-mixing angles and 1-Dirac phase. There is a clear mismatch between number of parameters on two sides and hence a one-to-one correspondence is impossible.
 Hence in case of HSMU for Majorana neutrinos one has to treat the Majorana phases as free parameters.  Since Majorana phases influence RG evolution of mixing angles, the predictions are subject to choice of Majorana phases. In case of
 HSMU for Dirac neutrinos the CKM and PMNS mixing parameters can be mapped in a one-to-one correspondence with each other at the unification scale. This leads to clear and unambiguous predictions.

 The RG running of masses and PMNS mixing  parameters for Dirac neutrinos can be found in \cite{Lindner:2005as}.   Like in previous section, we choose the unification scale = $2 \times 10^{16}$ GeV, SUSY breaking scale = 5 TeV and $\tan \beta$ = 55.  The dependence on these parameters will be discussed in Section \ref{sec4}.  Just like the Majorana case the HSMU hypothesis is implemented  in two steps. We start from  known values of gauge couplings, quark mixing angles, masses of quarks and charged leptons at low scale ($M_Z$). We then use the RG equations to obtain the corresponding values at high scale. At the high scale, in accordance with  HSMU hypothesis we take neutrino mixing angles and phase same as the quark mixing angles  and phase. The neutrino masses at high scale are taken as unknown parameters and we determine these three parameters such that the low energy values of the oscillation parameters i.e.  $\Delta m_{12}^2, \Delta m_{23}^2, \theta_{12}, \theta_{23}$ and $\theta_{13}$ agrees with 
their present experimental ranges.  The RG evolution of neutrino masses, mixing angles and Dirac phase are as shown in Fig. \ref{fig3}.

 \begin{figure}[ht]
            \vspace{0.2cm}
\begin{center}
\includegraphics[width=0.3\columnwidth]{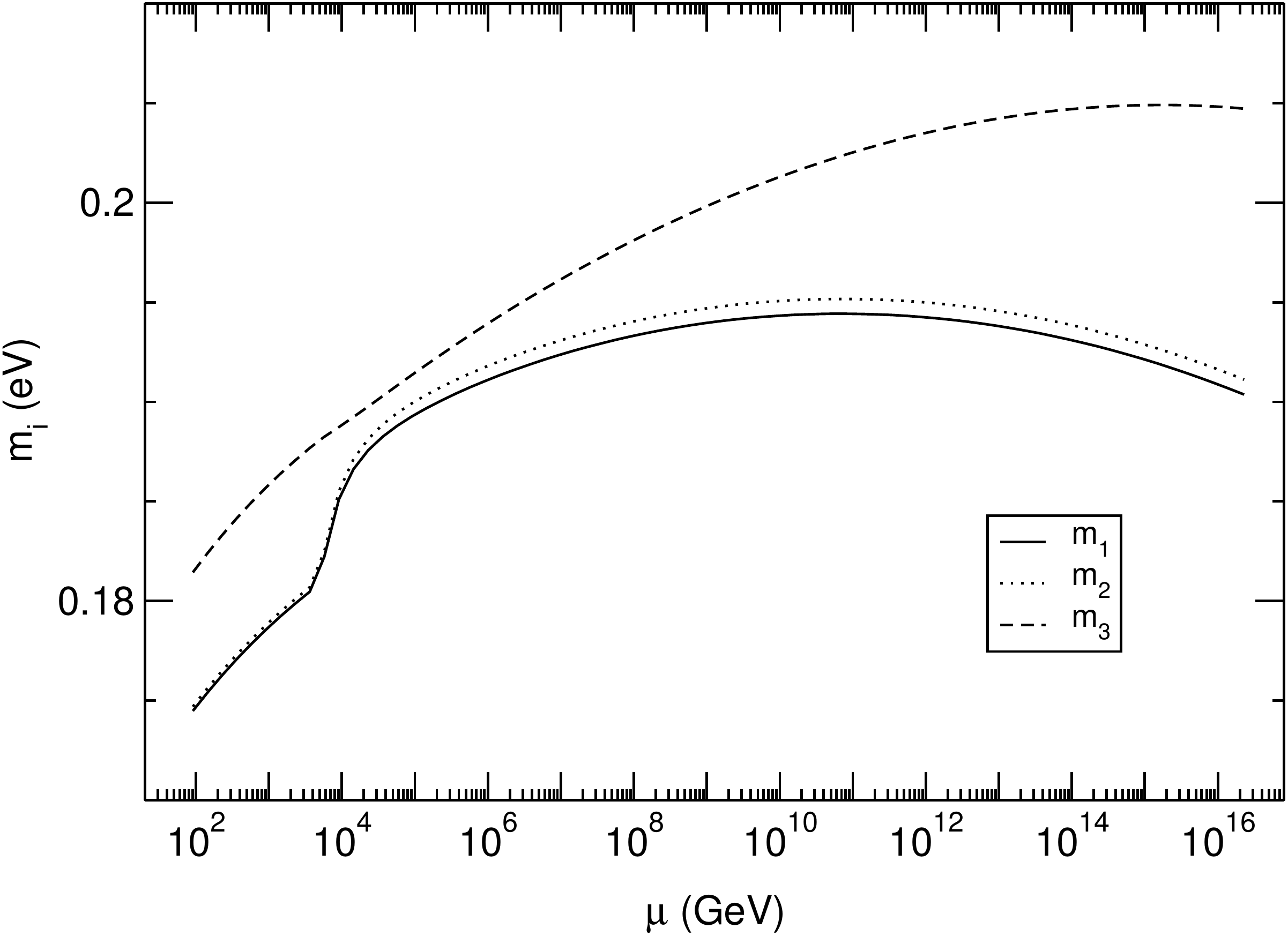} \hspace{0.1cm}
\includegraphics[width=0.3\columnwidth]{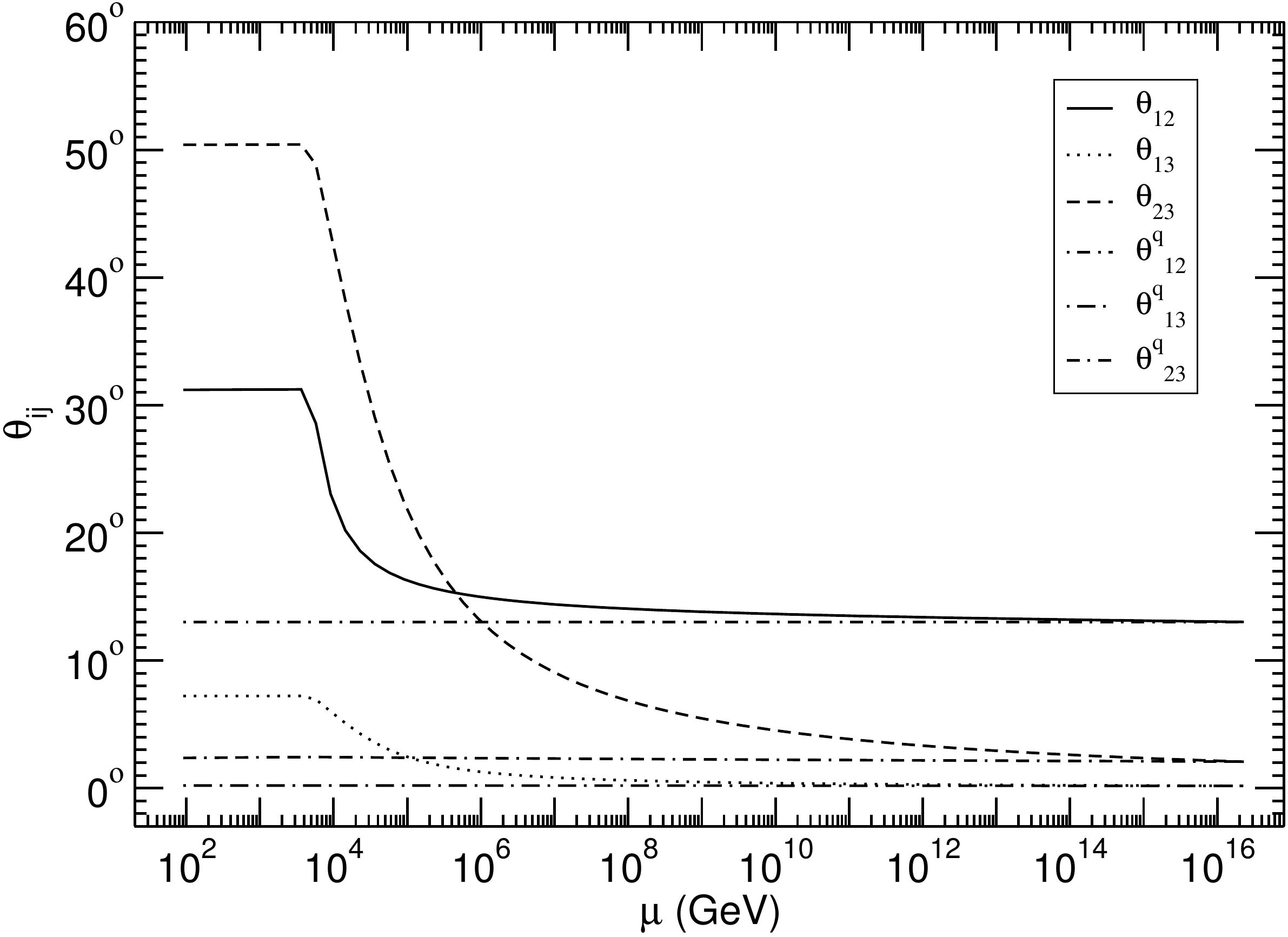} \hspace{0.1cm}
\includegraphics[width=0.3\columnwidth]{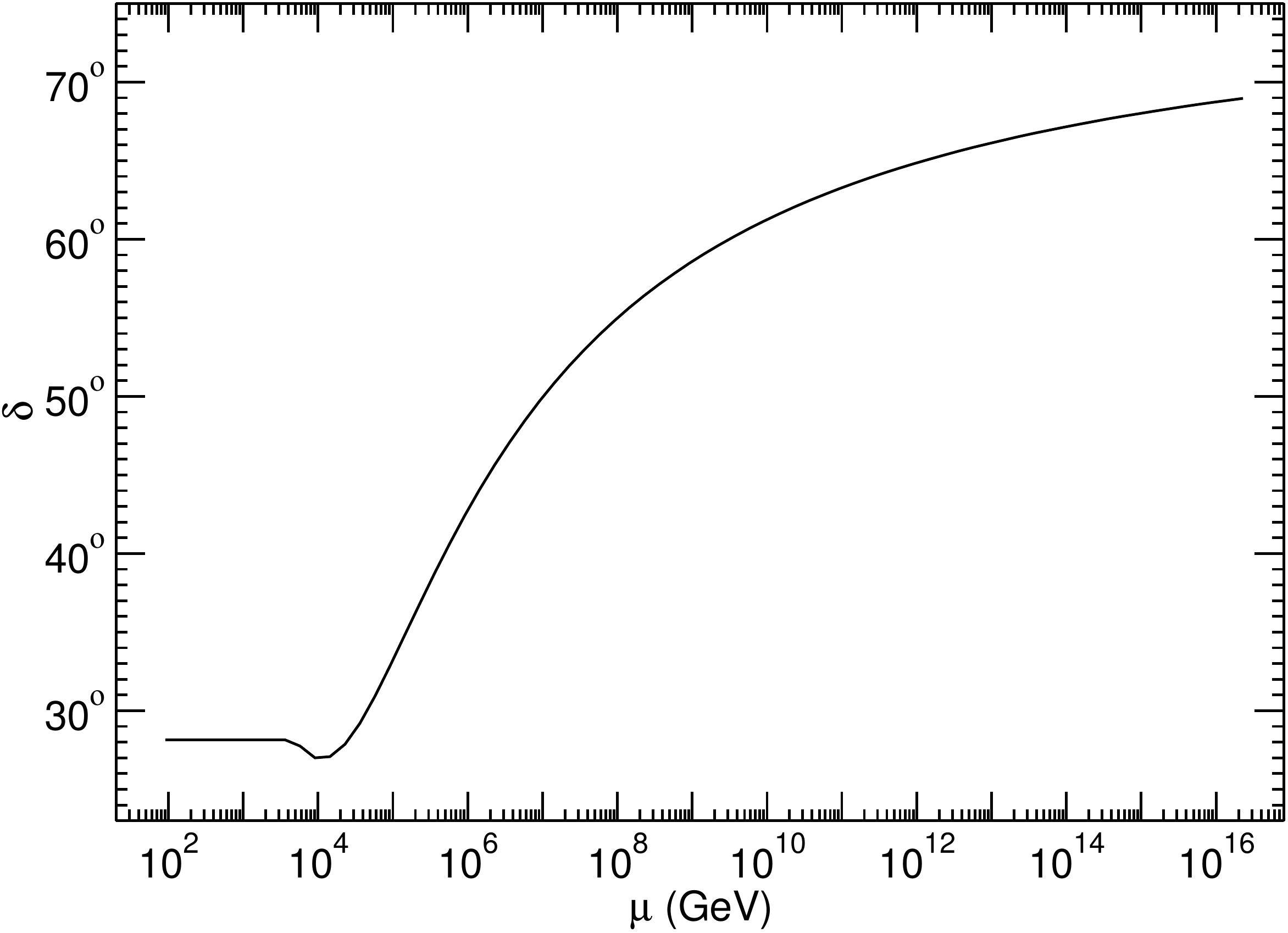}
\caption{The RG evolution of neutrino masses, CKM and PMNS mixing angles and Dirac phase.}
\label{fig3}
\end{center}
\end{figure}

\begin{figure}[ht]
   \vspace{0.2cm}
\begin{center}
\includegraphics[width=0.4\columnwidth]{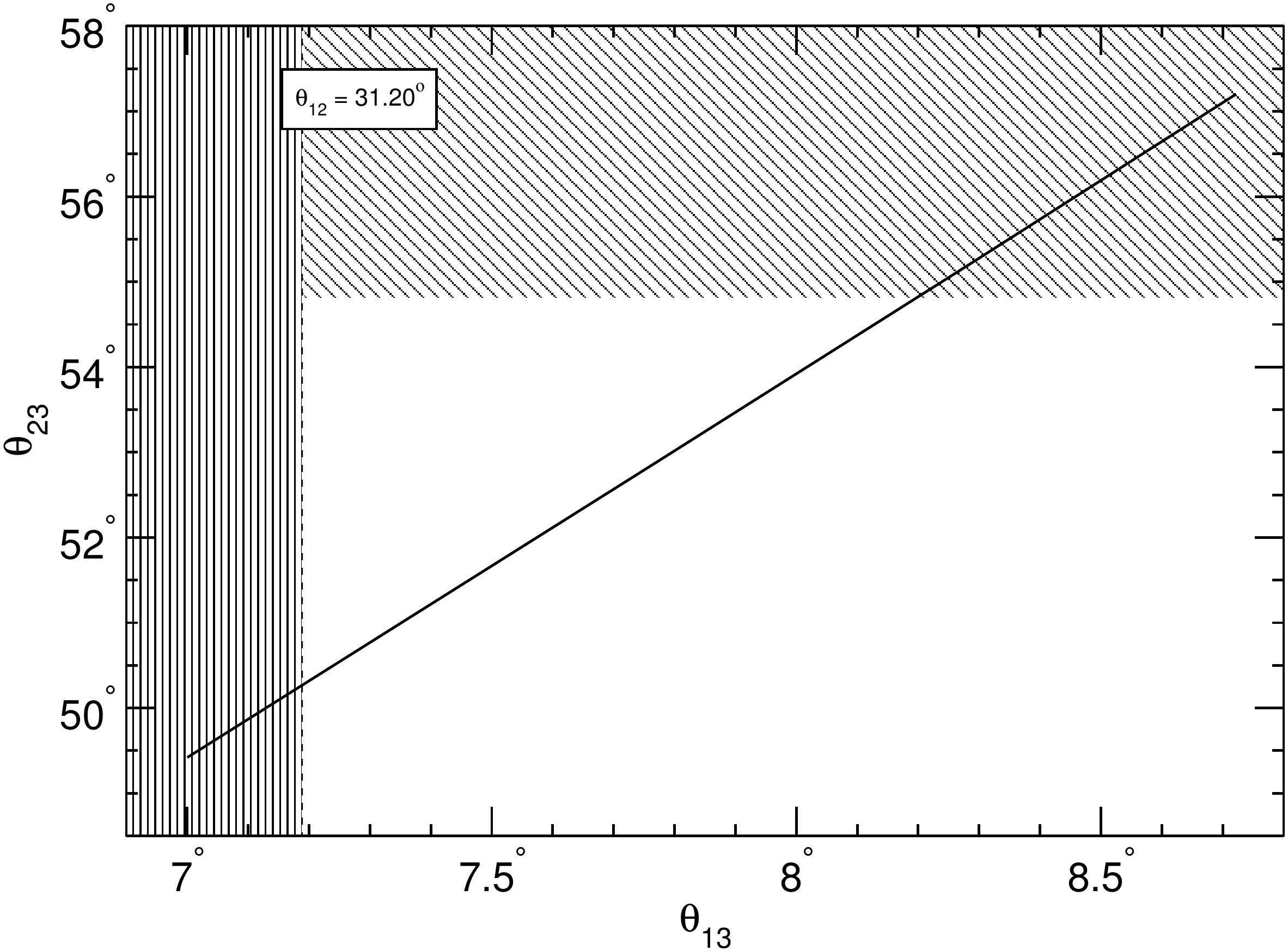}
\caption{The variation of $\theta_{23}$ with respect to $\theta_{13}$. The shaded regions lie outside the 3-$\sigma$ range \cite{GonzalezGarcia:2012sz}. }
\label{fig4}
\end{center}
\end{figure}

 As is clear from Fig. \ref{fig3}, owing to quasi-degenerate nature of neutrinos, large angle magnification occurs for the leptonic mixing angles.
  Since the RG evolution of $\theta_{13}$ and $\theta_{23}$ are correlated therefore as in previous case, here also $\theta_{23}$ turns out to be non maximal and lies in second octant as shown in Fig. \ref{fig4}.

After RG evolution all low scale parameters are within their 3-$\sigma$ range.  The mean mass of neutrinos at low scale is $m= 0.1769$ eV and the ``Averaged electron neutrino mass'' $m_\beta= 0.1747$ eV  which is slightly below the present reach of KATRIN experiment \cite{Drexlin:2013lha}. 
 
 Several important predictions can be derived for this case. The Dirac nature of neutrinos means that there is no neutrinoless double beta decay in this case. The ``Averaged electron neutrino mass'' $m_\beta$ turns out to be slightly below KATRIN's proposed sensitivity \cite{Drexlin:2013lha}. Since HSMU only works for normal hierarchy so that is another important prediction. Also, in this case  $\theta_{23}$ is non-maximal and always lies in second octant. There can be a small CP violation $\delta_{CP} \approx 15^\circ - 35^\circ $, $J_{\rm{CP}} \approx 0.1$.  These predictions provide important test of HSMU hypothesis for Dirac neutrinos and can be tested in present and near future experiments like INO, T2K, NO$\nu$A, LBNE, Hyper-K, PINGU \cite{Abe:2011ks,Patterson:2012zs,Adams:2013qkq,Ge:2013ffa,Kearns:2013lea,Athar:2006yb}.

%%%%%%%%%%%%%%%%%%%%%%%%%%%%%%%%%%%%%%%%%%%%%%%%%%%%%%%%%%%%%%%%%%%%%%%%%%%%%%%%%%%%%%%%%%%%%%%%%%%%%%%%%%%%%%%%%%%%%%%%%%%%%%%%%%%%%%%%%%%%%%%%%%%%%%%%%%%%%%%%%%%%%%%%%%%%

\section{Scale of HSMU and SUSY}
\label{sec4}

%%%%%%%%%%%%%%%%%%%%%%%%%%%%%%%%%%%%%%%%%%%%%%%%%%%%%%%%%%%%%%%%%%%%%%%%%%%%%%%%%%%%%%%%%%%%%%%%%%%%%%%%%%%%%%%%%%%%%%%%%%%%%%%%%%%%%%%%%%%%%%%%%%%%%%%%%%%%%%%%%%%%%%%%%%%%%%

 So far we assumed HSMU to be realized at GUT scale. but since HSMU does not depend on ``details'' of GUT scale theory it is instructive to analyze the effect of variation of HSMU scale. Similarly SUSY breaking scale and $\tan\beta$ were taken as 5 TeV and 55, respectively. It is important to analyze the dependence of HSMU on these. 
 
\begin{figure}[ht]
  \vspace{0.2cm}
\begin{center}
\includegraphics[width=0.3\columnwidth]{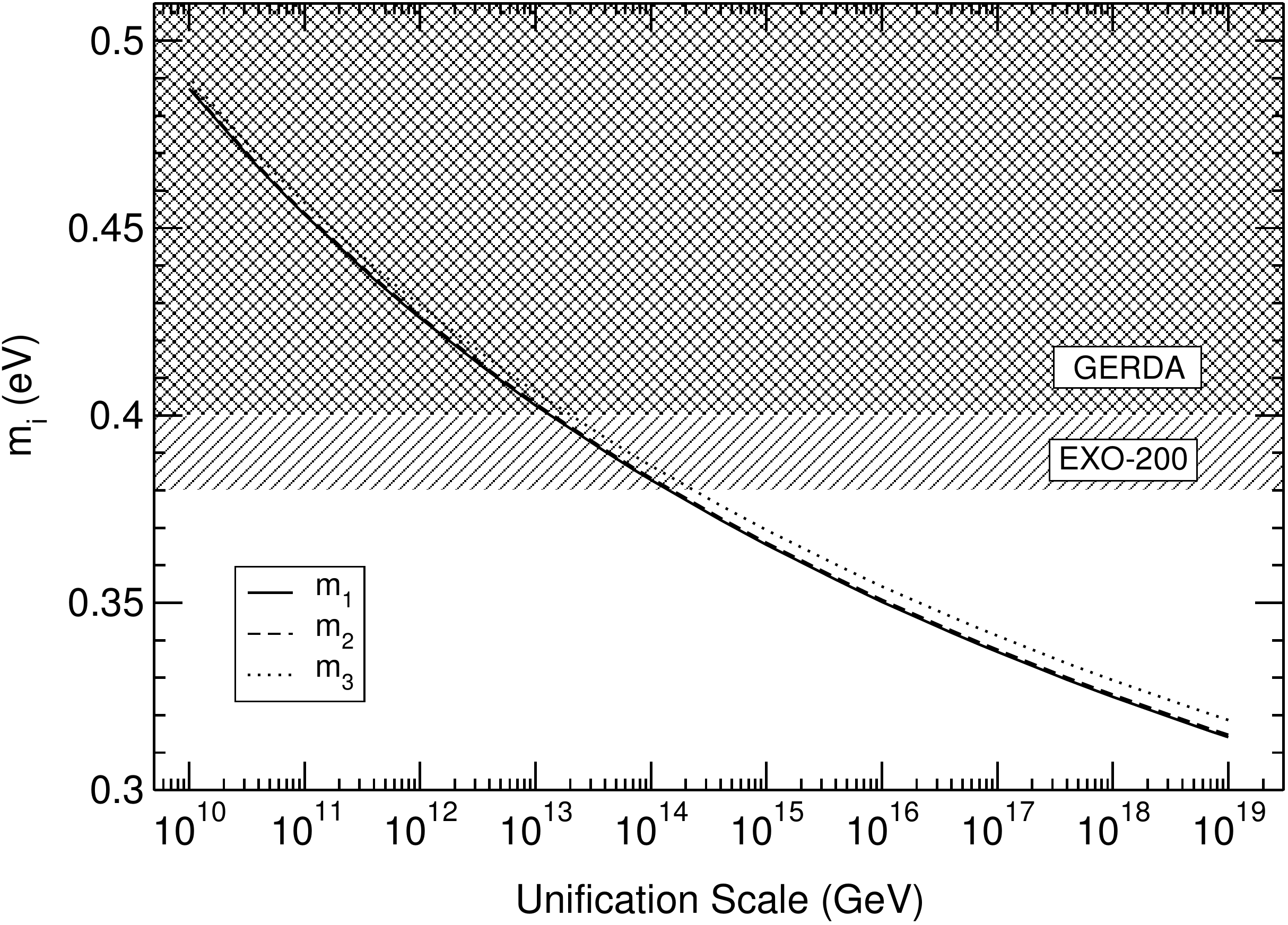} \hspace{0.1cm}
\includegraphics[width=0.3\columnwidth]{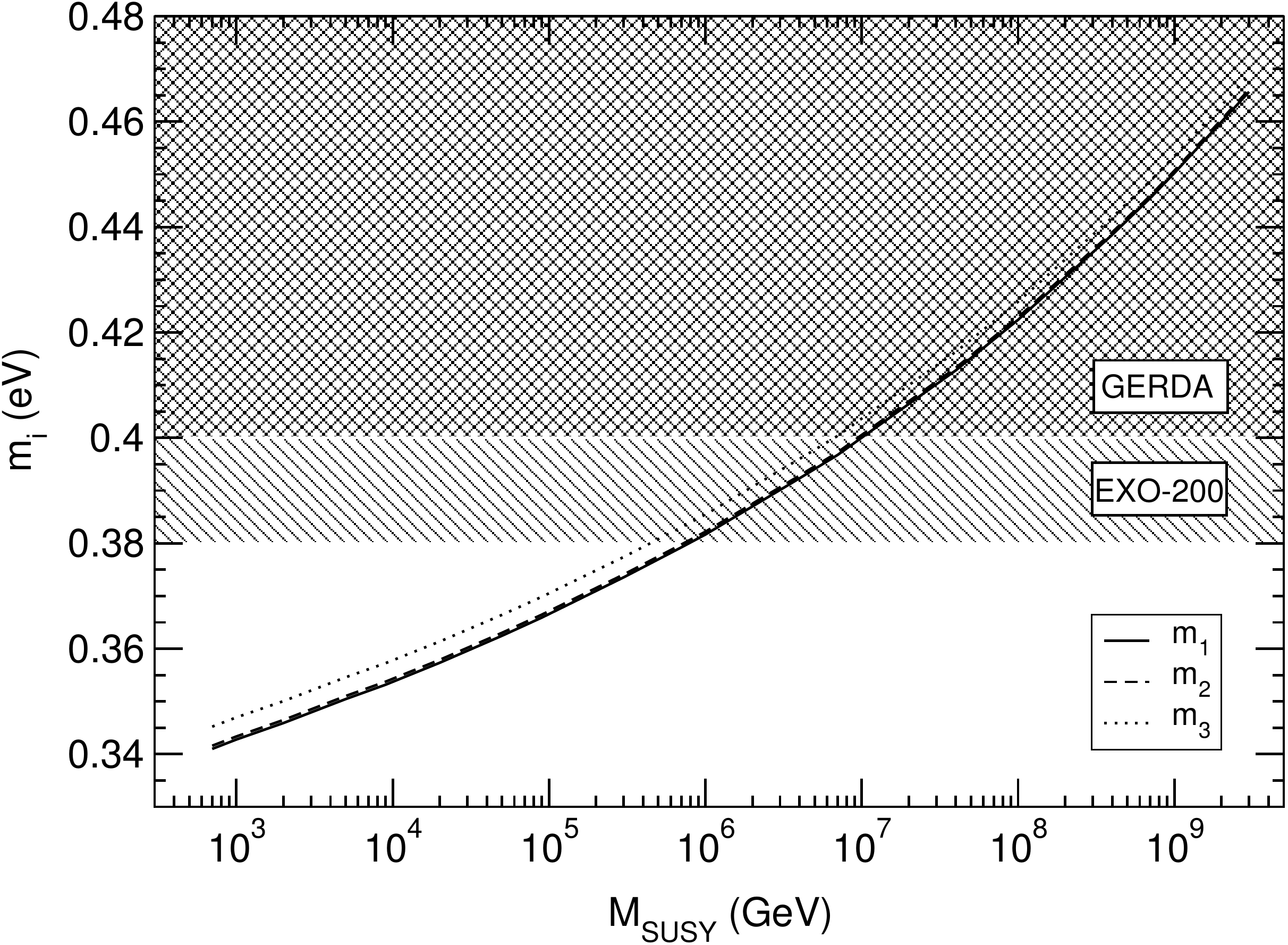} \hspace{0.1cm}
\includegraphics[width=0.3\columnwidth]{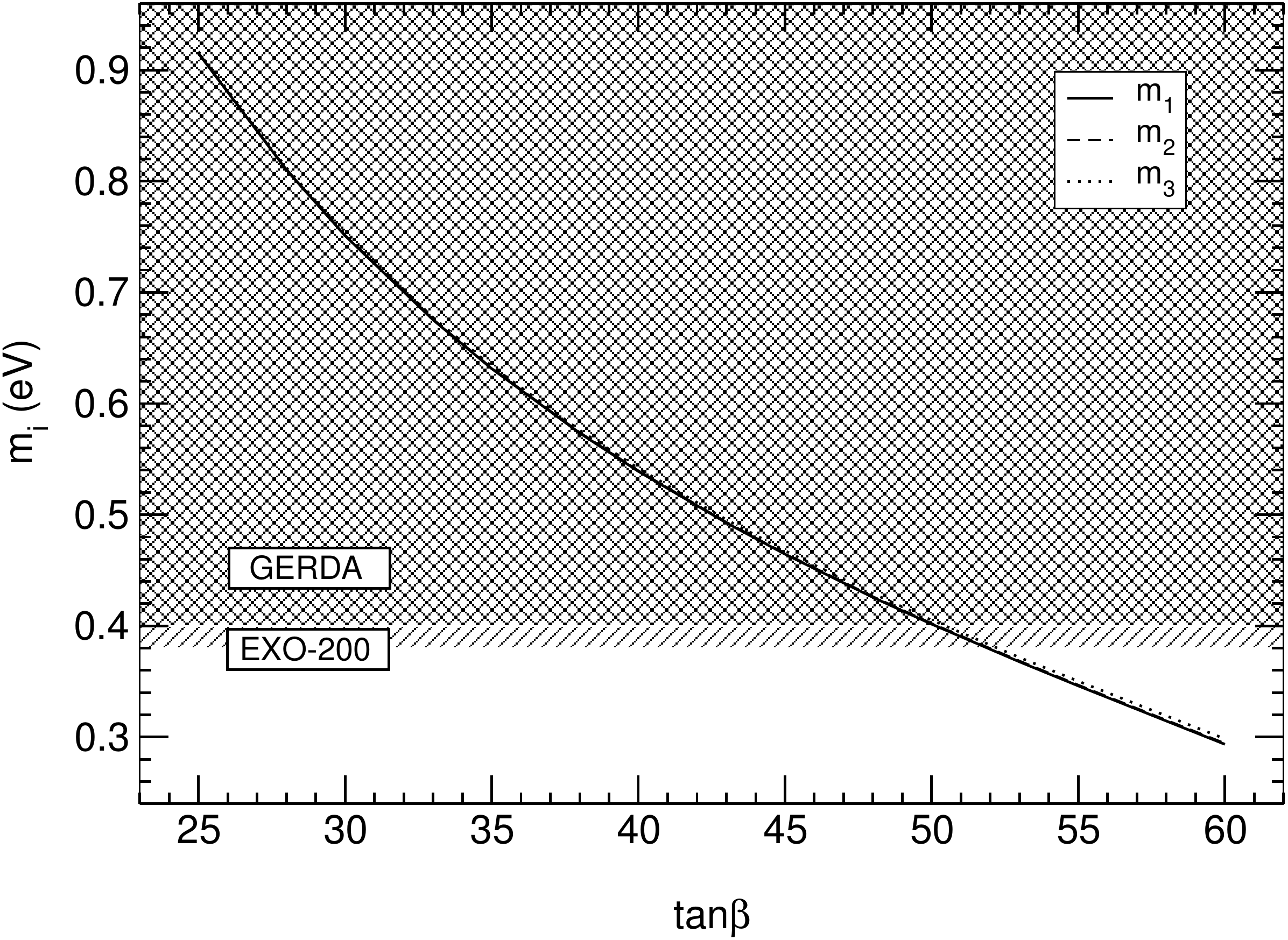}
\caption{The variation of $m_{\beta \beta}$ with respect to variation in unification scale, SUSY breaking scale and $\tan \beta$. }
\label{fig5}
\end{center}
\end{figure}

 Since the RG evolution of the mixing parameters and masses are correlated, the demand that all low scale oscillation parameters should lie within their 3-$\sigma$ range and all other observables like $m_{\beta \beta}$ should satisfy the current experimental bounds
 leads to very stringent constraints on the allowed values of unification scale, SUSY breaking scale and $\tan \beta$. In case of Majorana neutrinos the most stringent constraints on the unification scale, SUSY breaking scale and $\tan \beta$ are obtained from the upper bound on $m_{\beta \beta}$ as shown in Fig \ref{fig5}.

In plotting Fig \ref{fig5} we have kept all neutrino oscillation parameter at their best fit value.  The shaded regions in Fig \ref{fig5} are excluded by  $0\nu\beta\beta$ decay experiments \cite{Agostini:2013mzu,Auger:2012ar,Gando:2012zm,Alessandria:2011rc}. 
In the plot for variation of unification scale we have taken $M_{SUSY}$ = 5 TeV and $\tan \beta = 55$.  Similarly in the plot for variation of SUSY breaking scale we have taken unification scale = $2\times 10^{16}$ GeV and $\tan \beta = 55$ and in the plot for variation of $\tan \beta$ we have taken unification scale = $2\times 10^{16}$ GeV and SUSY breaking scale 5 TeV. As is clear from Fig \ref{fig5} experimental constraints require HSMU scale to be above $10^{13}$ GeV. Also SUSY breaking scales up to 1000 TeV are consistent with HSMU while only large values of $\tan \beta$ are allowed.

\begin{figure}[ht]
  \vspace{0.2cm}
\begin{center}
\includegraphics[width=0.4\columnwidth]{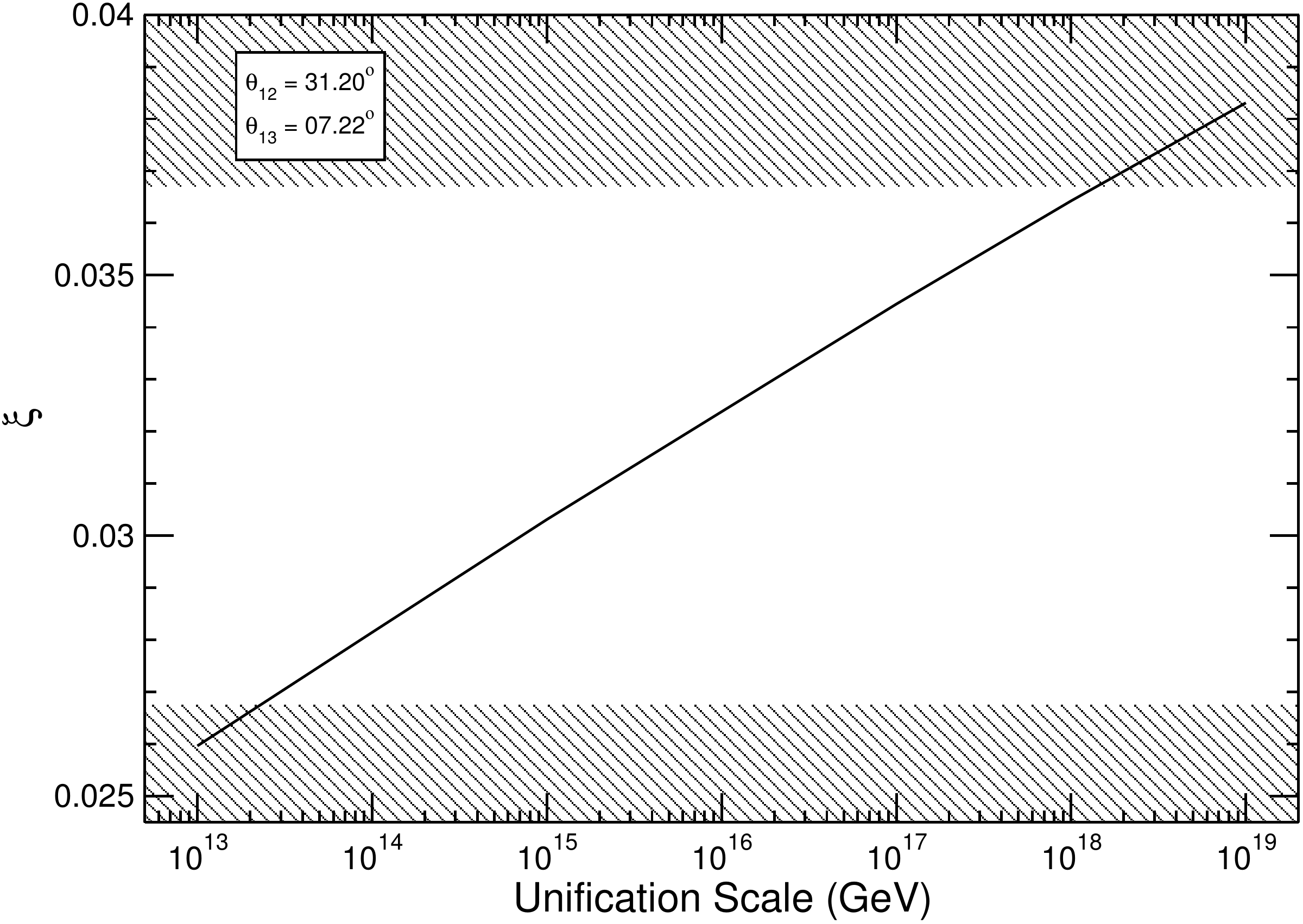}   \hspace{0.1cm}
\includegraphics[width=0.4\columnwidth]{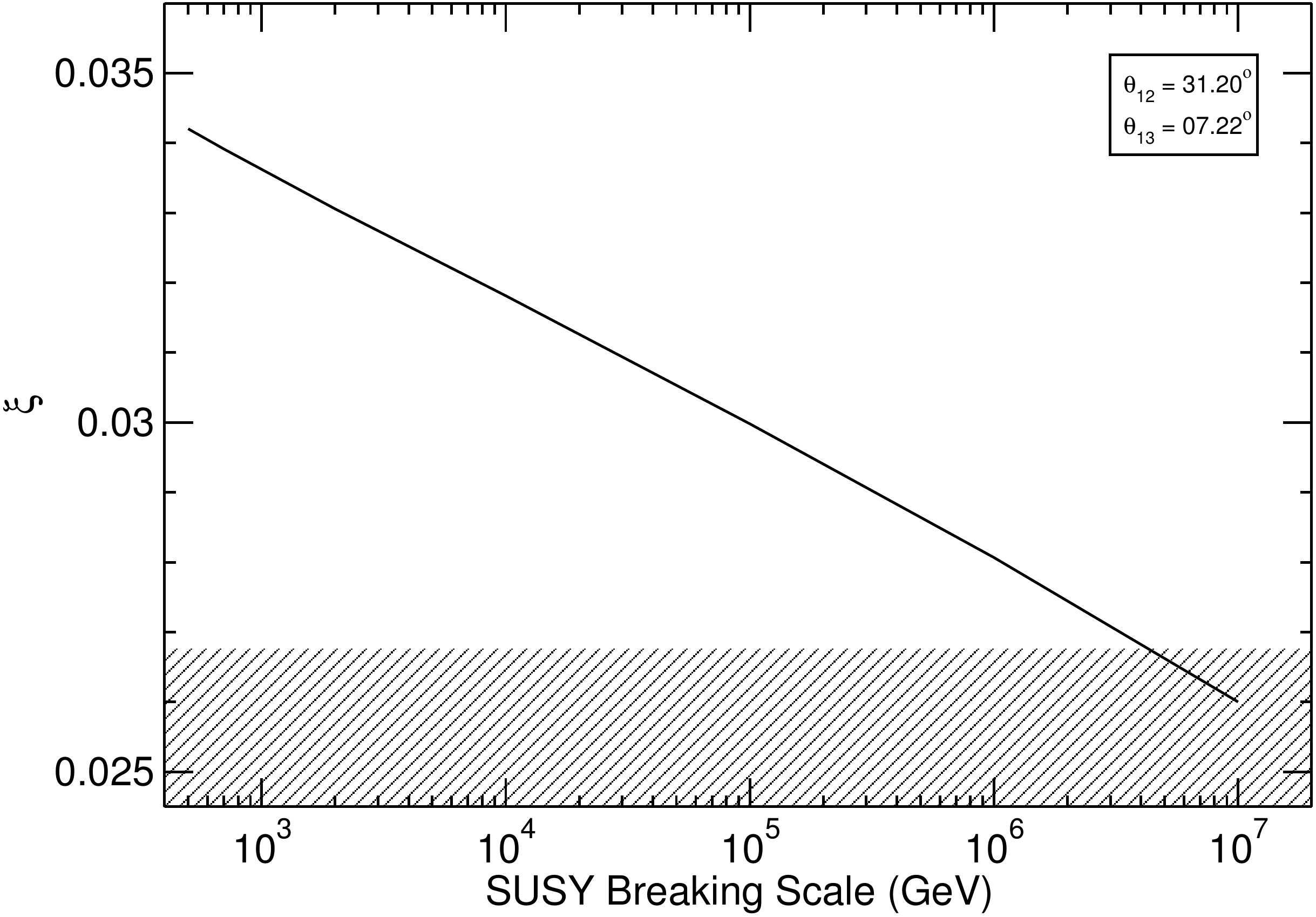}
\caption{The variation of $\xi$ with respect to variation in unification scale, SUSY breaking scale.}
\label{fig6}
\end{center}
\end{figure}

 For Dirac neutrinos also, similar constraints can be drawn. Since in case of Dirac neutrinos $m_{\beta \beta} = 0$ so it will not provide any constraint on the unification scale, SUSY breaking scale and $\tan \beta$. In this case the most stringent constraint are obtained from $\xi = \frac{\Delta m^2_{21}}{\Delta m^2_{32}}$ i.e. the ratio of the two mass square differences, as shown in Fig. \ref{fig6}.

The shaded regions in Fig \ref{fig6} are excluded by global fits of neutrino oscillation data \cite{GonzalezGarcia:2012sz}. In the plot for variation of unification scale we have taken $M_{SUSY}$ = 5 TeV and $\tan \beta = 55$.  Similarly in the plot for variation of SUSY breaking scale we have taken unification scale = $2\times 10^{16}$ GeV and $\tan \beta = 55$. As is clear from Fig \ref{fig6} experimental constraints require HSMU scale to be above $10^{13}$ GeV and SUSY breaking scales up to $10^7$ GeV are consistent with HSMU. Also the HSMU hypothesis is consistent with experimental constraints only for large values of $\tan \beta$.

%%%%%%%%%%%%%%%%%%%%%%%%%%%%%%%%%%%%%%%%%%%%%%%%%%%%%%%%%%%%%%%%%%%%%%%%%%%%%%%%%%%%%%%%%%%%%%%%%%%%%%%%%%%%%%%%%%%%%%%%%%%%%%%%%%%%%%%%%%%%%%%%%%%%%%%%%%%%%%%%%%%%%%%%%

\section{ Effect of Phases}
 \label{sec5}

% %%%%%%%%%%%%%%%%%%%%%%%%%%%%%%%%%%%%%%%%%%%%%%%%%%%%%%%%%%%%%%%%%%%%%%%%%%%%%%%%%%%%%%%%%%%%%%%%%%%%%%%%%%%%%%%%%%%%%%%%%%%%%%%%%%%%%%%%%%%%%%%%%%%%%%%%%%%%%%%%%%%%%%%%%%%%%

 In this section we look at the effect of the PMNS phases on the results derived in previous sections. As remarked earlier, in case of Majorana neutrinos HSMU cannot fix the Majorana phases at high scale. In Section \ref{sec2} we assumed no CP violation in leptonic sector. This assumption need not be realized in nature. The Dirac and Majorana phases (for Majorana neutrinos) enter RG equations of all other parameters and for certain choices they can have non trivial effects. Therefore, it is important to investigate their effects on the oscillation observables, in particular on the octant of $\theta_{23}$.
 
\begin{figure}[ht]
       \vspace{0.2cm}
\begin{center}
\includegraphics[width=0.4\columnwidth]{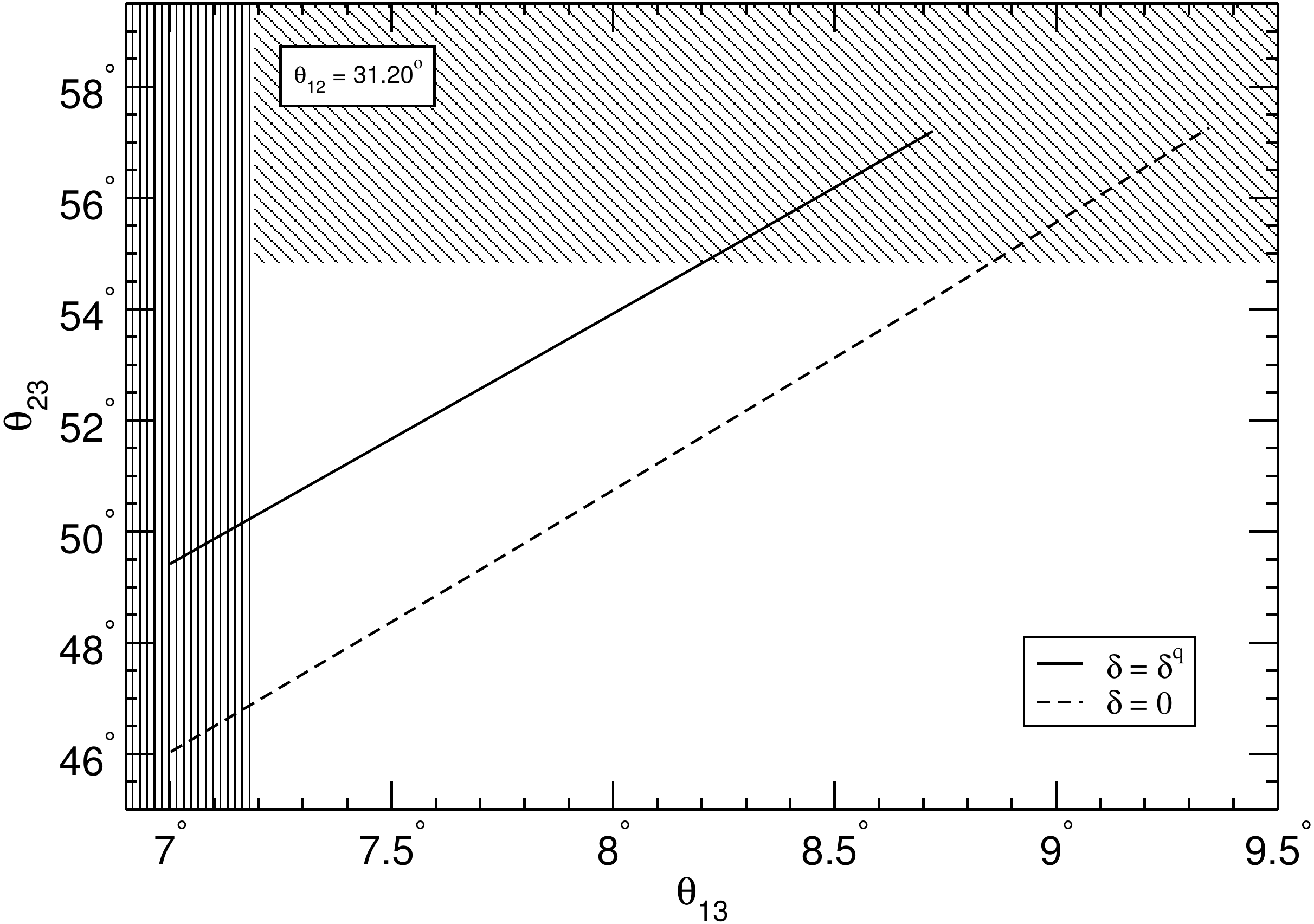}
\caption{The variation of $\theta_{23}$ with respect to $\theta_{13}$ for both CP conserving and violating scenarios. The shaded regions lie outside the 3-$\sigma$ range \cite{GonzalezGarcia:2012sz}.}
\label{fig7}
\end{center}
\end{figure}

 Before looking at the effect of phases on results obtained in Section \ref{sec2} for Majorana neutrinos, let us look at a simple possibility for the case of Dirac neutrinos.  In case of Dirac neutrinos, following HSMU hypothesis we took the PMNS Dirac phase to be same as CKM Dirac phase at high scales.  Another plausible scenario is the case of no CP violation in leptonic sector. For Dirac neutrinos if $\delta_{\rm{CP}}=0$ at high scale, it will remain zero at low scales.  RG effects can not regenerate  $\delta_{\rm{CP}}$ at low scales.  Such possibility will not change our conclusions obtained in Section \ref{sec3}. In particular $\theta_{23}$ will still remain non maximal and will always lie in second octant as shown in Fig. \ref{fig7}.  

\begin{figure}[ht]
  \vspace{0.2cm}  
\begin{center}
\includegraphics[width=0.4\columnwidth]{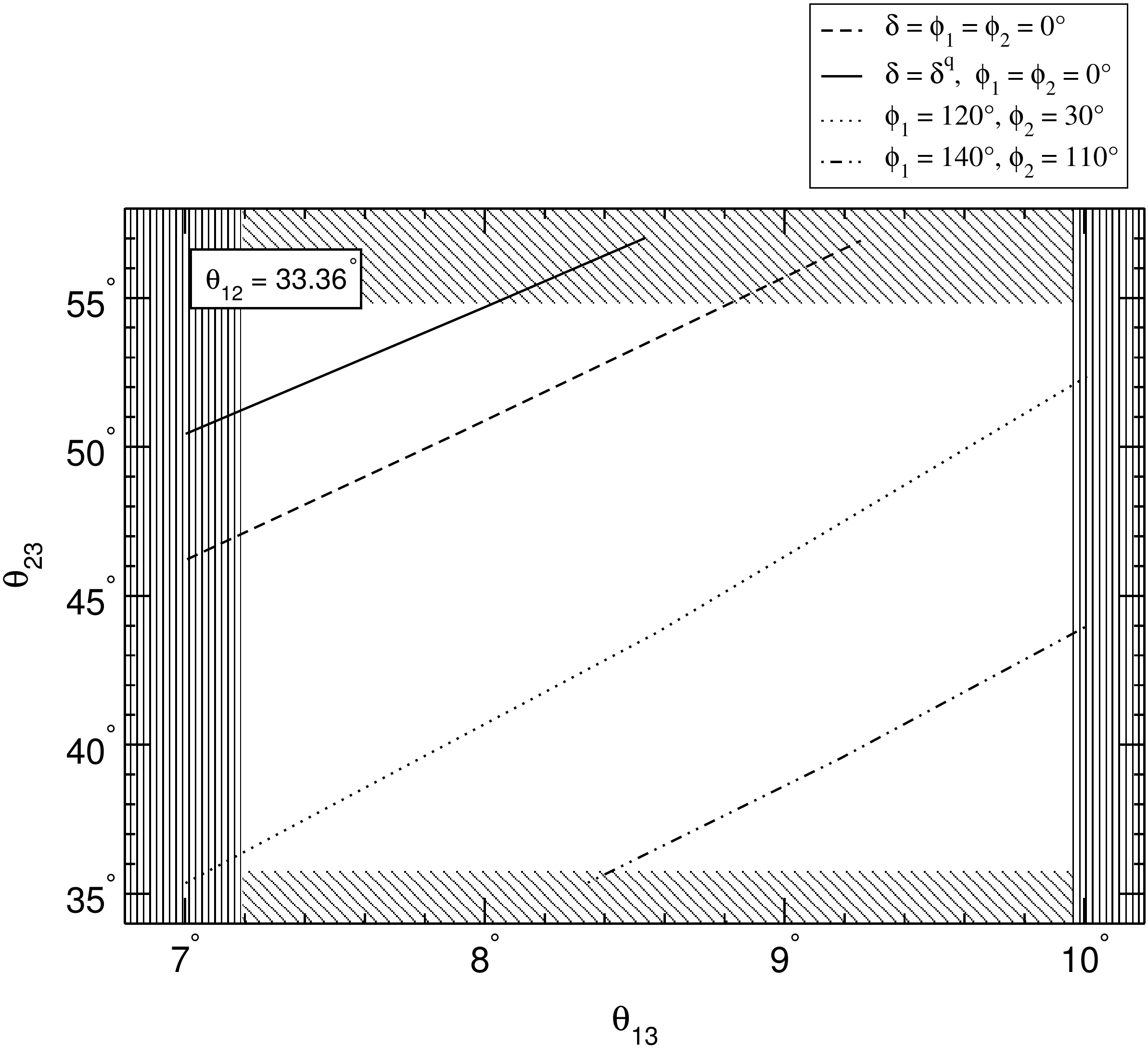}
\caption{The variation of $\theta_{23}$ with respect to $\theta_{13}$ for different choices of phases.. The shaded regions lie outside the 3-$\sigma$ range \cite{GonzalezGarcia:2012sz}.}
\label{fig8}
\end{center}
\end{figure}

In case of Majorana neutrinos the situation is more complicated and there are several possibilities. For example one simple possibility is that at high scale $\delta^0_{\rm{CP}} = \delta^{0,q}_{\rm{CP}} = 68.93^\circ $, $\phi_1 = \phi_2 = 0$. This implies CP violation in lepton sector at high scales.  But for $\phi_1 = \phi_2 = 0$ the RG running of $\delta_{\rm{CP}}$ results in a very small value at low scales.   In such scenario there is no CP violation at low scale and our conclusions will not change. In particular, $\theta_{23}$ will be non maximal and will always lie in second octant as shown in Fig \ref{fig8}. However the situation is different when we choose non-zero values of Majorana phases at the high scale. The phases have a ``damping effect'' on the RG evolution of the mixing angles and for certain choices they can lead to negligibly small magnification \cite{Agarwalla:2006dj, HSMU_phase}. In particular depending on the choice of phases, $\theta_{23}$ can lie in either octant and can also have 
maximal 
values as shown in Fig \ref{fig8}.

The presence of non trivial Majorana phases can have other consequences like appreciable CP violation in leptonic sector. Also, the phases effect the value of $m_{\beta \beta}$ which can now be as low as $0.2$ eV depending on the choice of phases. Also, the constraints obtained in Section \ref{sec4} on the allowed values of unification scale, SUSY breaking scale and $\tan \beta$ can be considerably relaxed. A detailed analysis on the effect of phases will be presented in \cite{HSMU_phase}. 

%%%%%%%%%%%%%%%%%%%%%%%%%%%%%%%%%%%%%%%%%%%%%%%%%%%%%%%%%%%%%%%%%%%%%%%%%%%%%%%%%%%%%%%%%%%%%%%%%%%%%%%%%%%%%%%%%%%%%%%%%%%%%%%%%%%%%%%%%%%%%%%%%%%%%%%%%%%%%%%%%%%%%%%%%%%%%

\section{Testing HSMU Hypothesis}
 \label{sec6}

%%%%%%%%%%%%%%%%%%%%%%%%%%%%%%%%%%%%%%%%%%%%%%%%%%%%%%%%%%%%%%%%%%%%%%%%%%%%%%%%%%%%%%%%%%%%%%%%%%%%%%%%%%%%%%%%%%%%%%%%%%%%%%%%%%%%%%%%%%%%%%%%%%%%%%%%%%%%%%%%%%%%%%%%%%%%%

The HSMU hypothesis is quite predictive and several currently running and near future experiments can test its predictions. We summarize our main results, along with the experimental tests that can be used to test our analysis in table below. 

 \begin{center}
\begin{tabular}{|c|c|c|}
  \hline
  Experiment                                    &    HSMU for  Majorana Neutrinos                         &    HSMU for Dirac Neutrinos                    \\
  \hline
  $m_{\beta\beta}$  (observed)                  &      Consistent                   &      Incompatible                 \\
 $m_{\beta\beta}<0.1$ eV                        &        Incompatible          &           Consistent                              \\
KATRIN $m_\beta$ (observed)                      &       Consistent            &          Consistent                                 \\
KATRIN $m_\beta$ (not observed)                  &         Incompatible          &           Consistent                                \\
$\theta_{23}>45^\circ$                          &       Consistent              &         Consistent                                  \\
  $\theta_{23}<45^\circ$                        &       Consistent            &           Incompatible                                  \\
Mass Hierarchy (Normal)                         &      Consistent              &         Consistent                                    \\
Mass Hierarchy (Inverted)                       &      Incompatible         &           Incompatible                                  \\
  \hline
 % \label{tab1}
  \end{tabular}
\end{center}
% \end{table}
% 

%%%%%%%%%%%%%%%%%%%%%%%%%%%%%%%%%%%%%%%%%%%%%%%%%%%%%%%%%%%%%%%%%%%%%%%%%%%%%%%%%%%%%%%%%%%%%%%%%%%%%%%%%%%%%%%%%%%%%%%%%%%%%%%%%%%%%%%%%%%%%%%%%%%%%%%%%%%%%%%%%%%%%%%%%%%%%

\section{Conclusion and Future Work}
\label{sec7}

%%%%%%%%%%%%%%%%%%%%%%%%%%%%%%%%%%%%%%%%%%%%%%%%%%%%%%%%%%%%%%%%%%%%%%%%%%%%%%%%%%%%%%%%%%%%%%%%%%%%%%%%%%%%%%%%%%%%%%%%%%%%%%%%%%%%%%%%%%%%%%%%%%%%%%%%%%%%%%%%%%%%%%%%%%%%%

  High Scale Mixing Unification of PMNS and CKM parameters is an interesting possibility.  As we have discussed it can be realized with both Dirac and Majorana type neutrinos. One of the important implications of HSMU is that it naturally leads to non zero and ``relatively large'' values of $\theta_{13}$ consistent with present global fits. Several predictions obtained from HSMU can be tested by present and near future experiments. Moreover, as we discussed the scale of HSMU is roughly same as that of Grand Unified 
  theories.  This opens up the possibility of realizing HSMU through a GUT.  Construction of such a GUT theory will put HSMU on a firmer footing and we are currently working towards the same.

%%%%%%%%%%%%%%%%%%%%%%%%%%%%%%%%%%%%%%%%%%%%%%%%%%%%%%%%%%%%%%%%%%%%%%%%%%%%%%%%%%%%%%%%%%%%%%%%%%%%%%%%%%%%%%%%%%%%%%%%%%%%%%%%%%%%%%%%%%%%%%%%%%%%%%%%%%%%%%%%%%%%%%%%%%%%%%%%%%%%%%%%%%%%%%%%%%%%%%%%%%%%%%%%%%%%%%%%%%%%%%%%%%%%%%%%%%%%%%%%%%%%%%%%%%%%%%%

\begin{acknowledgments}
I will like to thank the organizers for inviting me and giving me opportunity to present my work at the International Workshop on Unification and Cosmology after Higgs Discovery and BICEP2 - 2014. This contribution is based on \cite{Abbas:2013uqh, Abbas:2014ala, HSMU_phase} and I will like to thank G. Rajasekaran, G. Abbas and S. Gupta for a fruitful collaboration.  
\end{acknowledgments}

%%%%%%%%%%%%%%%%%%%%%%%%%%%%%%%%%%%%%%%%%%%%%%%%%%%%%%%%%%%%%%%%%%%%%%%%%%%%%%%%%%%%%%%%%%%%%%%%%%%%%%%%%%%%%%%%%%%%%%%%%%%%%%%%%%%%%%%%%%%%%%%%%%%%%%%%%%%%%%%%%%%%%%%%%%%%%%%%%%%%%%%%%%%%%%%%%%%%%%%%%%%%%%%%%%%%%%%%%%%%%%%%%%%%%%%%%%%%%%%%%%%%%%%%%%%%%%%

  %%%%%%%%%%%%%%%%%%%%%%%%%%%%%%%%%%%%%%%%%%%%%%%%%%%%%%%%%%%%%%%%%%%%%%%%%%%%%%%%%%%%%%%%%%%%%%%%%%%%%%%%%%%%%%%%%%%%%%%%%%%%%%%%%%%%%%%%%%%%%%%%%%%%%%%%%%%%%%%%%%%%%%%%%%%%%%%%%%%%%%

\end{document}